\documentstyle[floats,aps,prl,twocolumn,epsf]{revtex}

\newcommand{\vev}[1]{\langle #1 \rangle}

\newcommand{\prt}{\partial} 
 
\renewcommand{\v}[1]{{\bf #1}} 
\renewcommand{\t}[1]{{\tilde #1}}

\newcommand{\Del}{\Delta} 
\newcommand{\eps}{\varepsilon}

\newcommand{\om}{\omega}

\begin{document}

\twocolumn[\hsize\textwidth\columnwidth\hsize\csname
@twocolumnfalse\endcsname

\author{Patrick A. Lee}

\address{Department of Physics, Massachusetts Institute of Technology, Cambridge, MA 02139}

\title {\vspace*{-1cm}\hfill
{\it To be published in Proceedings of Euroconference on Complex Superconductors, Crete, Greece.}
       \vspace{1cm}\\Pseudogaps in Underdoped Cuprates}
\maketitle

\begin{abstract}
It has become clear in the past several years that the cuprates show many unusual properties,
both in the normal and superconducting states, especially in the underdoped region.  In
particular, gap-like behavior is observed in magnetic properties, $c$-axis conductivity and
photoemission, whereas in-plane transport properties are only slightly affected by the
pseudogap.  I shall argue that these experimental evidences must be viewed in the context of
the physics of a doped Mott insulator and that they support the notion of spin charge
separation.  I shall review recent theoretical developments, concentrating on studies based on
the $t$-$J$ model.  I shall describe a model based on quasiparticle excitations, which predicts
the doping dependence of $T_c$ and anomalous energy-gap--to--$T_c$ ratios.  Finally, I shall
outline how the model may be derived from a microscopic formulation of the $t$-$J$
model.  After a brief review of the $U(1)$ formulation, I shall explain some of the
difficulties encountered there, and how a new $SU(2)$ formulation can resolve some of the
difficulties.
\end{abstract}

\vspace{.25in}

]

\section{Introduction}
It has become clear in the past seveal years that the cuprates show many highly
unusual properties both in the normal and superconducting (SC) states.  These
unusual features are related to the fact that the cuprates are doped Mott
insulators.  It is then not surprising that the unusual behaviors are most
striking in the underdoped region, when the concentration of doped holes,
$x$ is small.   In the normal state a pseudogap is obseved in a temperature
range considerably above the SC transition temperature $T_{c}$.  The gap is seen
in NMR relaxation rate $1/T_{1}$, Knight shift \cite{Takigawa91} and specific
heat.\cite{Loram93}  It is also seen in $c$-axis conductivity \cite{Homes93} and in
photoemission experiments
\cite{Loeser,Ding}
which reveal that the pseudogap is roughly of the same size and $\v k$
 depdendence as the $d$-wave SC gap.  Furthermore, the gap size is
essentially independent of $x$ and even increases slightly when T$_{c}$ is reduced with
decreasing
$x$.  This observation is also supported by tunneling data.\cite{Renner98}
On the other hand, the in-plane transport properties are only slightly
affected by the pseudogap.  The resistivity shows a small decrease 
which may be interpreted as a decrease in the scattering rate.\cite{Puchkov}  
More
importantly, the spectral weight of the Drude  part of $\sigma
(\omega)$ is proportional to $x$ \cite{Orenstein} and there is no evidence that it is strongly
reduced by the presence of the pseudogap.\cite{Puchkov,Uchida,Tajima}  We believe this is strong
experimental evidence supporting the notion of spin-charge separation \cite{Anderson} in these
materials.  It was pointed out by P.W. Anderson \cite{Anderson} early on that the N\'{e}el state
is not the best way to accommodate the competition between the hole kinetic energy and the spin
exchange energy.  He envisioned another possibility, i.e., the spins form a liquid of singlets,
which he termed the resonating valence bond (RVB) state.  The reason is that the energy to form a
singlet $-J S(S+1)$ is particularly favorable for $S = \frac{1}{2}$.  The holes can more freely
among the liquid of singlets and are responsibile for the charge transport.  This notion of spin
charge separation naturally accounts for all the qualitative features of the spin gap state noted
above.  The spins form RVB singlets so that it costs energy (spin gap) to
make triplet excitations.  However the in-plane conductivity is carried by
$x$ holes, which remain gapless.  In $c$-axis conductivity and photoemission,
a physical electron is removed from the plane, which carries both spin and
charge.  It then follows that the spin gap should appear in these experiments.  This picture is
illustrated in Fig. 1.

\begin{figure}[tb]
\epsfxsize=2.85truein
\centerline{\epsffile{Pseudogaps.Fig.1}}
\vspace{0.25in}
\caption{ A cartoon representation of the RVB liquid
or singlets.  Solid bond represents a spin singlet
configuration and circle represents a vacancy.  In
Fig.~(1b) an electron is removed from the plane in
photoemission or $c$-axis conductivity experiment. 
This necessitates the breaking of a singlet.}
\end{figure}

We note that an alternative model which exhibits the above phenomenology is the model of
preformed pairs above $T_c$.  There are two versions of this class of model; the first
suggests that strong phase fluctuation \cite{Emery} destroys long range order over a large
temperature range above $T_c$, and the second assumes that we are in the short coherence
length limit of the pairing state \cite{Nozieres}, so that essentially pair ``molecules'' are
first formed and then Bose condensed.\cite{Uemura,de Melo}  The phase fluctuation model
predicts that the pairing amplitude is responsible for the pseudogap and would seem to
predict that other manifestations of superconductivity such as conductivity and diamagnetism
fluctuations should be observable, particularly at short distance and short time scale.  A
recent high frequency conductivity experiment in underdoped BISCO \cite{Corson} shows that
while Berenzinskii-Kosterlitz-Thouless (BKT) type fluctuations are observed near and above
$T_c$, the short distance (bare) superfluid density extracted from these measurements
vanishes above 100~K, much below the temperature range associated with the pseudogap.  These
data are difficult to understand within the phase fluctuation model.  Similarly, in the short
coherence length model, charge transport is by charge $2e$ pairs in the pseudogap state and
it is difficult to understand the insensitivity of the transport properties to the appearance
of the pseudogap with underdoping.  Furthermore, it is not at all clear that the coherence
length is short in the underdoped limit.  In section III we shall in fact argue that the
coherence length increases with underdoping, and that one is not in the short coherence length
regime.  In any event, in both these models a superconducting state with a large energy gap
is postulated to exist, without any indication of the origin and the energy scale of the
gap.  The RVB picture is fundamentally different from these preformed pair pictures in that
spin-charge separation plays a crucial role.  The pseudogap is a spin gap with an energy
scale set by $J$, which becomes the superconducting gap with the onset of coherence in the
charge degrees of freedom.  The superconducting state is characterized by spin-charge
recombination, forming superconducting quasiparticles which are quite conventional in the BCS
sense.

We model the cuprate with the $t$-$J$ model, which we believe contains the essential physics of
the doped Mott insulator.  The $t$-$J$ Hamiltonian is

\begin{displaymath}
H = \sum_{<ij>} J \left( {\bf S}_i \cdot {\bf S}_j - \frac{1}{4} n_in_j \right)
- t \sum_{\sigma} \left( c_{\sigma_i}^\dagger c_{\sigma_j} + h.c.   \right)
\end{displaymath}

\noindent
is subject to the constraint that double occupancy of a site by two electrons of opposite spins
is not allowed.  Here ${\bf S}_i = c_{i\alpha}^\dagger\mbox{\boldmath
$\sigma$}_{\alpha\beta}c_{i\beta}$ and $n_i = \sum_{\alpha}c_{i\alpha}^\dagger c_{i\alpha}$. 
The $t$-$J$ model is the strong coupling limit of the Hubbard model and the difficulty of its
solution lies in enforcing the no double occupancy constraint.  For the cuprates the
parameters are known to be $J \approx 0.13$~meV and $t/J \approx 3$.  When holes are doped
into the insulator, there is a gain in kinetic energy per hole proportional to $t$ due to
hopping.  However, the spin correlation is destroyed, costing an energy of approximately $J$
per site.  Thus we can consider the doping problem as a competition between the energy $xt$
(kinetic energy per site) and $J$.  When $xt << J$, the AF state with its doubled unit cell
is retained and the holes form small pockets around the top of the single hole dispersion,
which is known to be at $(\pi /2, \pm \pi /2)$ from photoemission \cite{Wells} (see Fig.~2).
This problem belongs to the same class as the doping
of a band insulator (or semiconductor).  The only
difference is that the coherent part of the band has
a reduced spectral weight of
$J/t$ and a bandwidth of order $J$.  This can be understood in terms of a spin-polaron
picture, i.e., the hopping hole is surrounded by a cloud of spin excitations.
\cite {Kane}  On the other hand, if $xt >> J$, the spin correlation becomes unimportant, AF
order is destroyed, and the holes should form a metallic state, describable by Fermi liquid
theory.  The Luttinger theorem then dictates that the area of the Fermi surface is given by
$1-x$, as shown in Fig.~2c.  The important point is the electrons which form the local moments in
the Mott insulator are now mobile and should be counted as part of the Fermi sea. The change in
Fermi surface area from $x$ to $1-x$ between the low and high doping limit is a special feature
of the doping of a Mott insulator, associated with the liberation of the local moments.  The
question then arises: how does the system evolve between these two limits?  The intermediate
state is apparently the spin gap state, with gaps in the one electron spectrum in the vicinity of
$(0,\pi)$ and segments of the Fermi surface near $(\pi /2,\pi /2)$ (see Fig.~2b).  As doping is
increased, these segments grow in length and eventually join to form the Luttinger Fermi
surface.  This intermediate state is clearly not a Fermi liquid because in band theory, gapping
of parts of the Fermi surface is not permitted without symmetry breaking.  However, the
breakdown of Fermi liquid theory is not a sharply posed issue at finite temperatures.  The
existence of the superconducting ground state at intermediate doping means that this question
cannot be investigated experimentally at present.  It is worth noting that the transition region
occurs near $x = 0.2$, when $xt$ and $J$ are comparable.

\begin{figure}[tb]
\epsfxsize=3.0truein
\centerline{\epsffile{Pseudogaps.Fig.2}}
\vspace{0.25in}
\caption{The evolution of the locus of low lying
single particle excitation with doping. (a) low
doping AF state: Brillouin zone is doubled with
small hole pockets of area
$x$.  (b) underdoped: Fermi surface ``segments.''
(c) overdoped: Fermi surface with area
$1-x$, satisfying Luttinger theorem.}
\end{figure} 

We emphasize once again that an important aspect of the doping of the Mott insulator is that the
resulting metal must remember that $x$ holes are responsible for the electrical conductivity.  If
the AC conductivity $\sigma (\omega)$ is characterized by a Drude-like component at low
frequency, we may characterize the conductivity by the scattering rate $1/\tau$ and the
spectral weight
$(n/m)_{\rm effective}$.  For underdoped samples, this spectral weight is proportional to
$x$.\cite{Orenstein}  This is very natural in that the weight must vanish when $x \rightarrow
0$.  It follows that a superconductor that forms out of the underdoped metal must have a
superfluid density
$\rho_s$ given by this spectral weight (in the clean limit), so that $\rho_s$ is proportional to
$x$.  This simple observation will play a prominent role in our subsequent discussion.  On the
other hand, when $xt > J$ we have a Fermi liquid state with electron density $1-x$.  The
question then arises as to how
$(n/m)_{\rm effective} \approx x$ can be accommodated within Fermi liquid theory.  Within Fermi
liquid theory we can write

\begin{equation}
\left( \frac{n}{m}\right)_{\rm effective} = \frac{1-x}{m^*} \left( 1 + \frac{F_{1S}}{2}
\right)
\end{equation}
\noindent
where $m^*$ is the effective mass and $F_{1S}$ is a Landau parameter.  It describes the
deviation of the current carried by the quasiparticle from $-e \v{v}_{\v{k}}$ due to dragging of
other quasiparticles

\begin{equation}
\v{j} = -e \alpha \v{v}_k
\end{equation}
\noindent
where $\alpha = (1 + F_{1S}/2)$.\cite{Millis}  Here we have made the simplifying assumption that
only the
$\ell = 1$ Landau parameter (in 2d) is important and the correction is independent of $\v{k}$. 
From Eq.~(1) we see that there are two ways to obtain $(n/m)_{eff} = x$.  The first is to
generate a heavy mass so that $m^* = 1/x$.  This is in fact the case for the system
La$_{1-x}$Sr$_x$TiO$_3$ which is a Mott insulator for $x = 0$ with a N\'{e}el temperature of
$T_N$ = 150~K.   With doping
$x > 0.02$, a metallic state is formed with the spin susceptibility $\chi$ and the specific heat
coefficient
$\gamma$ both scaling as $1/x$ and a Wilson ratio of order unity.\cite{Tokura}  The Hall
coefficient
$R_N
\sim 1/(1-x)$ as expected for a Fermi surface dictated by Luttinger theorem.  This is clearly a
realization of the Fermi liquid state expected for $xt > J$.  Unlike the cuprates, LaTiO$_3$ is a
three dimensional system, so that the exchange constant $J$ can be deduced from the ordering
temperature.  Thus the ratio
$J/t$ is very small and we believe this is the reason why the Fermi liquid state persists to low
doping.  For
$x < 0.02$, disorder effects become important and we are not able to explore the $xt < J$ limit
in this system.

A second route to achieve a spectral weight of $x$ is for $\left( 1 + \frac{F_{1S}}{2} \right)
\approx x$.  It turns out this is the route followed by the mean field slave boson theory described
below.\cite{Grilli}  We shall see that in the underdoped region this route is not followed in
the cuprate system.  We have strong evidence that the factor $\alpha$ in Eq.~(2) is not
proportional to
$x$ and is in fact of order unity.

\section{Microscopic Model and Mean Field Theory}
The physics of spin charge separation appears naturally in a class of theory
which starts with the $t$-$J$ model and enforces the constraint of no double
occupation by decomposing the electron into a fermion and a boson, $c_{i\sigma}^\dagger =
f_{i\sigma}^\dagger b_i$.  The fermion $f_{i\sigma}$ carries spin index and the boson $b_i$
keeps track of the charge degrees of freedom.  The constraint is replaced by the requirement
that
$f_{i\sigma}^{\dagger}f_{i\sigma} + b_{i}^{\dagger}b_{i} = 1$ which can be
enforced by introducing a Lagrangian multiplier so that field theoretic methods
may be applied.  This decomposition (called the slave boson method) is not
unique and one could just as well associate the spin with the boson (the
Schwinger boson theory\cite{Arovas}).  If the theories are solved exactly they
should give identical results.  However, different factorization leads naturally
to different approximation schemes.  Our strategy is to explore the different
schemes to see which correspond most closely with experiment.   In particular,
while the Schwinger boson method gives an excellent description of the
antiferromagnetic state at half filling,\cite{Arovas} it does not produce a large
Fermi surface for large doping.  Since we are mainly interested in the regime of
intermediate doping, the slave boson is a more promising starting point.

The exchange term can be written in terms of the fermions only\cite{Baskaran87}

\begin{eqnarray}
J\stackrel{\rightarrow}{S}_{i} \cdot \stackrel{\rightarrow}{S}_{j} & = & -J
\left| f_{i \alpha}^{\dagger} f_{j \alpha} \right|^{2} \nonumber \\
& = & -J \left(
f_{i\uparrow}^{\dagger}f_{j\downarrow}^{\dagger} -
f_{i\downarrow}^{\dagger}f_{j\uparrow}^{\dagger} \right) 
\left(
f_{i\downarrow}f_{j\uparrow} - f_{i\uparrow}f_{j\downarrow} \right)
\end{eqnarray} 

\noindent
which invites the following mean field decoupling

\begin{eqnarray}
\chi_{ij} & = & \langle f_{i\sigma}^\dagger f_{j\sigma} \rangle \;\;\;
\mbox{and} \nonumber \\ 
\Delta_{ij} & = & \langle f_{i\uparrow}f_{j\downarrow} - f_{i\downarrow}
f_{j\uparrow} \rangle 
\end{eqnarray}

\noindent
These parameters describe the  formation of a singlet on the bond $ij$.  The mean
field phase diagram\cite{Kotliar,Suzumura} is shown schematically in
Fig. 3.  As the temperature is lowered, $\chi_{ij} \neq 0$, so that the fermions
now acquire an energy band and a Fermi surface.  At a lower temperature, the
fermions form a pairing state with $d$-wave symmetry.  The bosons become
essentially Bose condensed (with exponentially large correlation length with
decreasing $T$) below a cross-over temperature $T_{BE}^{(0)} = 2 \pi xt$.  Below
$T_{BE}^{(0)}$ the boson field can be treated as a $c$-number.  In the overdoped
region this gives rise to a Fermi liquid phase, similar to the theory of heavy
fermion systems.  In the intermediate doping range, the simultaneous presence of
$\Delta_{ij}$ and $<b>$ gives rise to a pairing order parameter for physical
electrons $\langle c_{i\uparrow}c_{j\downarrow} \rangle$ which is of $d$-wave
symmetry.   Above $T_{BE}^{(0)}$ spin charge separation occurs at the mean field
level. In the pairing state a $d$-wave type gap occurs in the spin excitation
spectrum, but not in the charge excitation, and it is natural to identify this
as the spin gap phase.  Finally, the region IV in Fig. 3 is a non
Fermi liquid state which may be referred to as a ``strange metal.''

\begin{figure}[tb]
\epsfxsize=3.0truein
\centerline{\epsffile{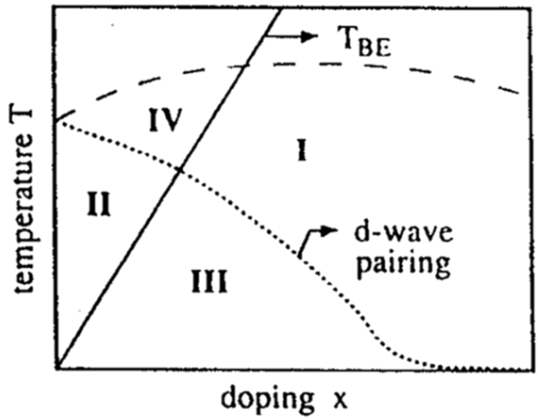}}
\vspace{0.25in}
\caption{Schematic mean-field phase diagram of the $t$-$J$ model.  Below the dashed line
the uniform RVB order parameter $\chi_{ij}$ is nonzero.  The mean-field pairing
line below which $\Delta_{ij} \neq 0$ (dotted) and the Bose-condensation line
(solid) divide the phase diagram into four regions.  Region I is a Fermi-liquid
phase, region II is the spin-gap phase, region III is the superconducting
phase, and region IV is the strange metal phase.}
\end{figure}

We can go beyond mean field to include fluctuations about the mean field
solution.  The most important fluctuations are the phase fluctuations of the
order parameter $\chi_{ij} = | \chi_{ij} | e^{i\theta_{ij}}$.  Particles
hopping around a plaquette acquire a phase related to $\theta_{ij}$, just like
electrons in the presence of a magnetic flux.  These low lying excitations are
U(1) gauge fields.\cite{Baskaran} We shall refer to this theory as the $U(1)$ formulation. When
coupled to the fermions and bosons they enforce the constraint locally, not just on average as in
mean field theory.

\section{Phenomenological Description of the Superconducting State}
Before continuing with the microscopic theory, we digress to review a phenomenological description
of the superconducting state.\cite{LeeWen}  The idea is to start at low temperature where the nature
of the elementary excitations is well understood, and calculate the reduction of the superfluid
density with increasing temperature.  This can be done by making two assumptions: A)~the  superfluid
density is given by $x$, and  B)~the quasiparticle (qp) dispersion in presence of an external
electromagnetic gauge potential has a BCS form:
\begin{equation}
 E^{(sc)}_{\v A}(\v k)=\pm \sqrt{\eps^2(\v k)+
 \Delta^2(\v k) }+ \frac{\v A}{c} \cdot \v j (\v k)
 \label{Esc}
\end{equation}
for $\v k$ near the nodes, where $\v{j}(\v{k})$ is given by Eq.~(2). Note that  
$\v v_F \equiv \v \prt_{\v k} \eps$ is the ``normal state'' Fermi velocity and 
that the vector potential $\v A$ couples only through
the ``normal state dispersion'' $\eps(\v k)$ and has nothing to do with 
the SC gap $\Del(\v k)$.  The physical reason for this is that this quasiparticle is a
superposition of an electron with momentum $\v k$ and a hole with momentum $-\v k$, and both these
objects carry the same charge current $\v j(\v k)$.  Mathematically Eq.~(5) is easily derived
by noting that $\v A$ enters only in the diagonal elements of the BCS matrix in the form
$\epsilon (\v k + \v A)-\mu$ and $-\epsilon (-\v k + \v A)-\mu$, which is diagonalized to
give Eq.~(5) to linear order in $\v A$.

With these assumptions we can calculate how the superfluid density is reduced by
thermal excitation of quasiparticles.  We found that

\begin{equation}
\frac{\rho_{s}}{m}(T) = \frac{x}{a^{2}m}-\frac{2\alpha^2\ln 2}{\pi} \frac{v_{F}}{v_{2}} T
\end{equation}

\noindent
where $v_{2} = \Delta_{0} a/\sqrt{2}$ is the velocity of the quasiparticle in the
direction of the maximum gap $\Delta_{0}$, i.e., in the direction from the node towards
$(0,\pi)$. The ratio $v_{F}/v_{2}$ thus measures
the anisotropy of the massless Dirac cone which characterizes the $d$-wave qp
spectrum.

Based on numerical calculations and theoretical considerations, \cite{Dagotto,Laughlin} we
expect the mass
$m$ in Eq.~(6) to correspond to a tight-binding hopping integral of order $t$ so that $m^{-1}
\sim t$.  Experimentally it is found that $m$ is about twice the electron mass
\cite{Orenstein}, which happens to correspsond to a hopping integral of $J \approx 0.13$~eV.
We believe that the theoretical expression for the hopping integral is $t/3$ which happens to
equal $J$ in our case.  On the other hand, the Fermi velocity $v_{F}$ is proportional to the
coherent bandwidth, which is given by $J$.  To keep our expression general, we keep track of
the distinction between $t/3$ and $J$, even though numerically they are equal.

We see that for small $x$, the quasiparticle excitation is an effective way of destroying the
superconducting state by deriving $\rho_s$ to zero.  By extrapolating Eq.~(6) to
$\rho_s = 0$, we can estimate $T_c$ as being of order 

\begin{equation}
T_c \approx x\Delta_0(t/3J) \,\,\,\, .
\end{equation}
\noindent
Note that this is a violation of the BCS ratio $2\Delta_0/kT_c = {\rm constant}$.
Presumably the real transition is driven by critical fluctuations including phase
fluctuations and vortex unbinding in the 2d limit, but the underlying (bare) superfluid
density should be driven to zero by quasiparticle excitations in the way we indicate.  There
is experimental support for this point of view from high frequency conductivity
measurements.\cite{Corson}  If we further assume that $\Delta_0$ is independent of $x$ for
underdoped cuprates, we see that $T_c$ is proportional to $x$ (or more precisely to $\rho_s
(T=0)/m$), thus providing an explanation of Uemura's plot. \cite{Uemura}  

It is worth noting
that in the slave boson mean field theory $\Delta_0$ is proportional to $J$, since that is
the only energy scale relevant to the formation of spin singlets.  Then Eq.~(7) predicts that
$T_c$ is proportional to $xt$.  Apart from numerical coefficients, this has the same
functional dependence as the Bose condensation temperature $T_{BE}^{(0)}$, as well as the
transition temperature based on pairing of bosons to be discussed later in the $SU(2)$
formulation. 

Another important implication is that superconductivity is destroyed when only a small fraction of
the quasiparticles (with energy $\leq x \Delta_0$) are thermally excited.  Thus the gap near
$(0,\pi)$ must remain intact in the normal state, leaving a strip of thermal excitations which
extend a distance proportional to $x$ from the nodal points.  This is qualitatively in agreement
with the photoemission experiment.  Of course our phenomenological picture does not provide a
description of the normal state.  It simply states that the normal state gap is an inescapable
consequence of a finite $\Delta_0$ and a vanishingly small superfluid density as $x \rightarrow
0$.

The fact that $d\rho_s/dT$ is independent of $x$ and that both $\rho_s$ and $T_c$ are proportional
to $x$ means that a scaled plot of $\rho_s(T)/\rho_s(0)$ vs $T/T_c$ should be independent of $x$
for small $T/T_c$.  In fact, such a scaled plot for YBCO$_{6.95}$ and YBCO$_{6.60}$ shows a
remarkable universality over the entire temperature range.\cite{Bonn}  
We can use the data to extract
the ratio $\alpha^2v_F/v_2$ using Eq.~(4).  Using the YBCO$_{6.95}$ data, we obtain a
velocity anisotropy $v_F/v_2 = 6.8$ if we assume that $\alpha = 1$.\cite{LeeWen}  

Alternatively, by comparing the measured slope $d\rho_s/dT$ of the YBCO$_{6.95}$ and YBCO$_{6.60}$
samples, we see that the slopes are almost the same, showing that $\alpha^2v_F/v_2$ is almost
independent of doping.  From tunneling data we know that the maximum gap $\Delta_0$ slightly
increases with underdoping.\cite{Renner98} This implies that $\alpha$ is almost independent
of
$x$.  This is the experimental evidence that the Fermi liquid scenario $\alpha = x$ does not
apply to the underdoped cuprates.

It is useful to compare Eq.~(4) with the standard BCS expression which is usually written in the
form \cite{Hirschfeld}

\begin{mathletters}
\begin{eqnarray}
\rho_s(T) & = & \rho_s(0) \left( 1- \frac{(2 \ln 2)T}{\Delta_0} \right)  \\
          & = & \rho_s(0) - \rho_s(0) (2 \ln 2) T/\Delta_0
\end{eqnarray}
\end{mathletters}
\noindent
This expression does not include Fermi liquid correction and should be compared with Eq.~(6)
with
$\alpha = 1$.  We note that in BCS theory $\rho_s(0)$ is independent of $x$ and the second term in
Eq.~(8) is in exact agreement with the second term in Eq.~(6) as it should be, because the
derivation leading to Eq.~(6) is completely general.  The first terms in Eq.~(8) and Eq.~(6)
do not agree because the standard BCS theory does not apply to a doped Mott insulator and
does not include the physics leading to a spectral weight proportional to $x$.  It is clear
that this feature of Eq.~(8) does not agree with experiment on underdoped cuprates.  If one
ignores this and fits the normalized data $\rho_s(T)/\rho_s(0)$ to Eq.~(8a), one would reach
the incorrect conclusion that the energy gap $\Delta_0$ is proportional to $T_c$ in
underdoped cuprates.
\cite{Panagopoulos}  We emphasize that Eq.~(6) includes Eq.~(8) as a special case and must be
used in place of Eq.~(8) for a correct analysis of the data.

We can also estimate the size of the vortex core using this picture.  The idea is to identify the
core size as the point where the critical current is reached.  If we replace $-e \v{A}/c$ in
Eq.~(2) by the gauge invariant superfluid velocity 
$
\v{v}_s = \frac{1}{2} \left( \v{\nabla}\theta - \frac{2e}{c} \v{A}\right)
$,
we see that the quasiparticle energy shifts up or down in the presence of $v_s$ and quasiparticles
are generated at the Fermi energy, contributing to a normal fluid density.  Near the vortex core,
$v_s$ grows as $1/R$, so that the normal fluid density grows and eventually drives the critical
current to zero.  This allows us to estimate the core size to be

\begin{equation}
R_1 \approx \frac{1}{x} \frac{v_F}{\pi \Delta_0(t/3J)} \approx \frac{v_F}{\pi T_C}
\end{equation}

\noindent
Note the factor $x$ appears in the denominator.  We note that in BCS theory, the coherence length
can be written either as $v_F/\pi T_c$ or $v_F/\Delta_0$.  The two forms are equivalent because the
ratio $2\Delta_0/kT_c$ is a constant.  In our case this ratio depends on $x$ and it is not clear
{\it a priori} which form is correct for the coherence length.  Equation~(9) shows that
$v_F/\pi T_c$ is the correct form for the coherence, and not $v_F/\pi \Delta_0$.  One
consequence of this is that the underdoped cuprates are in fact not short coherence length
superconductors.\cite{de Melo}  The number of holes per coherence volume actually grows as
$x^{-1}$ with decreasing doping.  A second consequence is that $H_{c2}$ (due to orbital
effects) is predicted to scale as $x^2$.  Within this picture it is also clear that in
underdoped cuprates the state inside the vortex core should retain the large gap $\Delta_0$,
just as the normal state above $T_c$.

We can now estimate the condensation energy using the relation $\Delta E = H_c^2/8\pi$ and $H_c^2 =
H_{c1}H_{c2}$.  Noting that $H_{c1}$ is proportional to $\rho_s(0)/m \approx xt$ while $H_{c2} =
\phi_0/R_1^2$ is proportional to
$x^2$, we find that $\Delta E$ is proportional to $x^3$, i.e.,

\begin{equation}
\Delta E \approx x(T_c^2/J)(t/3J)
\end{equation}

\noindent
This is in contrast to the BCS expression $\Delta E \approx T_C^2/\epsilon_F$.  Equation~(10)
also follows from a picture where only the quasiparticles with energy less than $T_c$ are
affected by the transition to the normal state.  The area of the Brillouin zone occupied by
these excitations is of order
$(T_c/J)(T_c/\Delta_0)$, so the total energy change per area is of order $T_c^3/J\Delta_0$
which agrees with Eq.~(10).  Thus even when expressed in terms of $T_c$ the condensation
energy is much less than the BCS value in the underdoped system.  There is evidence for this
suppression of the condensation energy from specific measurements.\cite{Loram93}

 \section{The $\v{SU(2)}$ Formulation of the ${\v t}$-${\v J}$ Model}
We now return to discuss the microscopic theory.  While the mean field phase diagram is in
qualitative agreement with experiments, the $U(1)$ formulation suffers from a number of
deficiences if we try to improve the mean field theory by including gauge fluctuations at the
Gaussian level.  In the spin gap phase the problem lies with the fact that the MF theory is a
pairing theory of fermions and carriers with it some features of superconductivity.  For
example, the gauge field is gapped by the fermion pairing via the Anderson-Higgs mechanism. 
This leads to a reduction of gauge fluctuations which actually destabilize the pairing phase.
\cite{Ubbens}  A second problem is that if we introduce residual interaction between the
fermions and bosons to form an electron, the electron spectrum will always have nodes.  This
is because the node structure in the pairing state is tied to the Fermi level and is very
resilient to interactions.  Thus we have difficulty reproducing the ``Fermi surface
segments'' which are apparently observed in photoemission experiments.  In the
superconducting phase we have condensation of the bosons and the quasiparticles become well
defined.  While this feature is in agreement with experiment, the current carried by the
quasiparticles turns out to be reduced so that in Eq.~(2), $\alpha = x$.  As we have seen,
this leads to a serious disagreement with the doping dependence of the temperature
coefficient of the London penetration depth.  In order to circumvent these difficulties, we
were led to a new formulation of the $t$-$J$ model which is designed to be more accurate near
half filling.  We briefly outline the $SU(2)$ formulation below.\cite{Wen,LNNW}

In this new formulation we introduce an $SU(2)$ doublet of boson
fields $b^{T}=(b_{1},b_{2})$, in addition to the fermion doublet
$\psi^{\dagger}=(\psi_{\uparrow},\psi_{\downarrow}^{\dagger})$.  The physical
electron is represented by the $SU(2)$ singlet formed out of these two
doublets, $c_{\uparrow}=\frac{1}{\sqrt{2}}b^{T}\psi$,
$c_{\downarrow}=\frac{1}{\sqrt{2}}b^{\dagger}\bar{\psi}$ where
$\bar{\psi}=i\tau^{2}\psi^*$.  We are motivated by the observation made by
Affleck $et$ $al$.\cite{Affleck} that at half-filling $(x=0)$ the fermion
representation of the $t$-$J$ model has the $SU(2)$ symmetry in that a spin-up
electron can be represented by a spin-up fermion or the absence of a spin-down
fermion.  In the $U(1)$ formulation this symmetry is broken as soon as $x \neq 0$,
and out of a infinte degeneracy of states, the $d$-wave fermion pairing state is
picked out as the MF solution.  In contrast, even at the mean field level, the low
lying states which are missing in the $U(1)$ mean field theory are included in the new $SU(2)$
formulation.  For example, the spin gap state can be described equally well as the $d$-wave
pairing state, or a staggered flux phase, where the fermions see gauge fluxes
which alternate from plaquette to plaquette.  The $SU(2)$ gauge
transformation relates these states and guarantees that there is no breaking of
the translational symmetry.  The fermion spectrum
exhibits a $d$-wave type gap, with maximum gap at $(\pi, 0)$ and nodes at
$(\pi/2, \pi/2)$.  We compute the physical electron spectral function, which at
the mean field level, is a convolution between the fermion and boson spectra. 
We further introduced a residual interaction between the fermions and bosons. 
The resulting spectra can be compared with photoemission experiments and have the following
features.  The spectra consist of a coherent part with spectral weight $x$ and dispersion of
order $J$ and a broad incoherent part.  The coherent part closely resembles the fermion
dispersion.  The residual interaction broadens and shifts the nodes at $(\pi/2, \pi/2)$ so
that we obtain a ``Fermi surface segment'' near
$(\pi/2, \pi/2)$.  Away from this segment a gap appears in the excitation
spectrum which grows to its maximal magnitude near $(0,\pi)$.  This behavior is
in qualitative agreement with the angle-resolved photoemission
experiment.\cite{Loeser,Ding}

We have also studied the fermion spectrum and how it is affected by gauge fluctuations.  We found a
logarithmic correction to the fermion velocity and we successfully fitted the magnetic
susceptibility and the specific heat in the spin gap state.\cite{Kim}

In the superconducting state we need to address the issue of the current carried by the
quasiparticles.  To expand on this point further, we note that in the original $U(1)$ gauge
field formulation of the
$t$-$J$ model, the prediction for $\rho_s(T)$ takes the form of Eq. (6) with $\alpha = x$ and
therefore is in strong disagreement with experiment.  This follows simply from the Ioffe-Larkin rule
which states that the inverse of the response function of the fermion and boson should add to give
the physical inverse response.  In the superconducting state, the fermion and boson acquire
superfluid densities $\rho_F$ and $\rho_s$ so that

\begin{equation}
\rho_s^{-1}(T) = \rho_F^{-1}(T) + \rho_B^{-1}(T)
\end{equation}
where $\rho_F \approx (1-x)$ and $\rho_B \approx x$.  However, only the temperature dependence of
$\rho_F$ depends on the qp gap structure and is expected to be of the form $\rho_F(T) \approx
(1-x)(1-T/\Delta_0)$, whereas the temperature dependence of $\rho_B$ arises only through the
excitation of sound mode and should be higher power in $T$, which can be ignored.  Inserting these
into Eq. (11) we see that $\rho_s(T)$ is predicted to be $x - x^2 T/\Delta_0$.    Basically in
the
$U(1)$ gauge theory the mismatch of the Fermi surface area and the Drude spectral weight (or
$\rho_s$ in the superconducting state) is solved by a Landau parameter, so that $\alpha =
x$.  Thus we may conclude that it is not sufficient to treat the gauge fluctuation only to
quadratic order as in the Ioffe-Larkin theory.

We believe this difficulty is tied to the notion of Bose condensation as a way of achieving
superconductivity.  The reason is the following.  The electron operator $c_k$ is a convolution of
the fermion and boson operator in momentum space.  Let us suppose that the external $\v A$ field
couples only to the boson (this is true in the $SU(2)$ formulation and is approximately true
in some gauge choice  in the $U(1)$ formulation).  In the presence of $\v A$, $b_\v q
\rightarrow b_{\v q +
\v A}$ so that after the convolution $c_\v k \rightarrow c_{\v k + \v A}$ and $\epsilon_\v k
\rightarrow \epsilon_{\v k + \v A}$ as expected.  Thus $j_\v k = -e \partial \epsilon / \partial \v
A = -e \partial \epsilon / \partial \v k$.  Let us see what happens in the superconducting state. 
If we assume that the fermions are already paired, superconductivity can be driven by the
condensation of bosons $<b_{\v k = 0}> = 0$.  However, in the presence of $\v A$, the Bßose
condensate remains rigid and stays in the $k = 0$ state.  This is clearly seen in the Ginsburg
Landau theory for the free energy 
$|\left( \v \nabla - 2e \v A /c \right) b|^2$
where $<b_{\v k=0}> \neq 0$ in the presence of $\v A$ is responsible for the Higgs mechanism and the
London penetration depth.  Upon convolution, we see that for the electron operator, $\v k$ is not
shifted by $\v A$ so that $\epsilon(\v k)$ is independent of $\v A$.  The qp now carries no
current!  In the $U(1)$ formulation, the gauge field $\v a$ causes a small shift in the
Fermion spectrum and leads to Eq. (2) with $\alpha = x$.  This is clearly an unacceptable
situation and can be seen most acutely for the qp at the Fermi surface along the ($\pi$,$\pi$)
direction.  Here the energy gap vanishes so that the qp in the superconducting state is
basically the same state above
$T_c$.  Yet, according to the Bose condensation scenario, the current carried by this qp drops
abruptly below $T_c$.

Now that we have identified the problem, we can see that there are two possible ways to avoid it. 
The first is to argue that due to fluctuations, only a small fraction of the bosons are in the
condensate and we can reduce the problem, but not eliminate it.  We call this the single boson
condensation (SBC) scenario.  The result is that $\alpha$ can lie anywhere between $x$ and 1, and
most likely somewhere in between. A second possibility is allowed in the $SU(2)$ formulatin
but not in the $U(1)$ formulation.  In $SU(2)$ theory there are two species of bosons $b_1$
and
$b_2$ and we can pair them to form a gauge singlet pair
$<b_1 (\v i) b_2(\v j)> \neq 0$.  We shall call this the boson pair condensation (BPC) scenario. 
Since $<b_1> = <b_2> = 0$, the problem is avoided and we find that $\alpha = 1$.  This is really a
consequence of continuity because in this scenario the superconducting qp along ($\pi$,$\pi$) is
smoothly connected to the electron state above $T_c$.  This result comes out of an explicit
calculation which we outline below.\cite{WenLee} 

In $SU(2)$ theory we go beyond MF theory by calculating 
the electron propagator
through a ladder diagram \cite{Wen,LNNW} to include effects of pairing
between the boson and the fermion. Here we will consider only the simplest
on-site interaction
$V\left(c^\dagger _\uparrow c_\uparrow + c^\dagger _\downarrow c_\downarrow
\right) $, 
which, when written in terms of bosons and fermions, generates an attraction
 between bosons and the fermions if $V>0$.
There are also other pairing interactions, but they will not modify our
results qualitatively. The resulting electron propagator is given by
\begin{eqnarray}
&& \v G_{\v A}(\om,\v k) \equiv
 \pmatrix{ -i \vev{ c_\uparrow c_\uparrow^\dagger} & -i\vev{c_\uparrow c_\downarrow} \cr
  -i \vev{ c_\downarrow^\dagger c_\uparrow^\dagger} & -i\vev{c_\downarrow^\dagger c_\downarrow} \cr}
\\
 &=& \left[ 
 \pmatrix{ 
 G_{0,\v A}(\om,\v k) &  F_{0,\v A}( \om, \v k) \cr
 F_{0,\v A}(\om,\v k) & -G_{0,\v A}(-\om,-\v k) \cr
 }^{-1} - V\tau^3 \right]^{-1}
 \label{mc15}
\end{eqnarray}

We first consider the second scenario 
where there are no SBC,  but there is a nonzero $F_{0,\v A}$ proportional to the boson pair
parameter $x_{pc}$. For $\v A=0$, the poles of $G_{11}(\om, \v k)$ comes in pairs of opposite signs,
just as in BCS theory. However the total residue is $\frac{x}{2(1-VG_{in})^2}$, significantly reduced
from the BCS value. There are two positive branches which determine the
qp excitations
\begin{equation}
 E^{(sc)}_\pm (\v k)= \sqrt{ \t E_\pm^2 + \left(\frac{x_{pc}}{x} \Del \right)^2}
 \label{mc16}
\end{equation}
where
\begin{equation}
 \t E_\pm = \pm \sqrt{(\eps-\t \mu)^2+\Del^2
   -\left(\frac{x_{pc}}{x} \Del \right)^2} -\t \mu 
 \label{mc17}
\end{equation}
and $\t \mu =-\frac{xV}{4(1-VG_{in})}$. 
In order to interpret those results, let us
first consider the normal state which is recovered by setting $x_{pc}=0$ in
Eq. (14) and Eq. (15), yielding the normal state dispersion
$E^N_\pm \equiv \t E_\pm(x_{pc}=0)$. This corresponds to
a massless Dirac cone initially centered
at $(\pm \pi/2, \pm \pi/2)$ when $V=0$ which is the MF fermion spectrum
of the staggered-Flux ($s$-Flux) 
phase. The effect of $V$ (the boson-fermion pairing) is
two-fold. The $\t \mu$ inside the square-root shift the location of the node
towards $(0,0)$ by a distance $\Del k=-\t \mu/v_F$ while the last term shift
the spectrum upwards. The cone intersects the Fermi energy to form a small
Fermi pocket with linear dimension of order $x$.
As shown in Fig. 4(a), the spectral weight is concentrated on one side of the
cone, so that only a segment of FS on the side close to the origin
carries substantial weight.  This is the origin of the notion of ``FS segment''
introduced in Ref. \cite{Wen,LNNW}.

\begin{figure}[tb]
\epsfxsize=3.75truein
\centerline{\epsffile{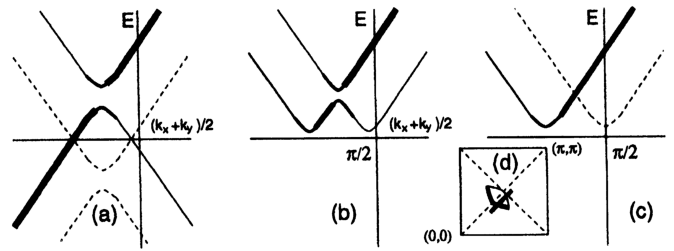}}
\vspace{0.25in}
\caption{Schematic illustration of the QP dispersion (the pole location of {\bf G} for
(a) normal state, and SC state with (b) $0 < x_{pc} < x$ and (c) $x_{pc} = x$.  The line
thickness indicates the size of the residue of $G_{11}$, and the dashed line indicates
vanishing residue.  The momentum scan is along the straight line in (d), where the curved
segment is the FS segment in the normal state.}
\end{figure}

Now let us see what happens in the SC state when $x_{pc}\neq 0$. Equation (14) takes
the standard BCS form if $\t E_\pm$ is interpreted as the normal state
dispersion. However, $\t E_\pm$ differs from the normal state spectrum
$E_\pm^N$ by the appearance of the term $-(x_{pc}\Del/x)^2$ in Eq. (15).
Close to the node this term is small so that qualitatively the spectrum
develops from the normal state in a BCS fashion, as shown in Fig. 4(b). This is
particularly true if the higher energy gap between the two branches is smeared
by lifetime effects. Thus we see that the ``FS segment'' is gapped
in a BCS-like fashion. However, the velocity $v_2$ in the $(1,-1)$ direction,
being proportional to $x_{pc}/x$, does not extrapolate to the gap at
$(0,\pi)$ (which is essentially independent of $x_{pc}$), but crosses over to it
at the edge of the FS segment. It is worth remarking that in the
special case $x_{pc}=x$, $E^{(sc)}_\pm$  reduces to the standard BCS form
with the normal state dispersion $\eps(\v k)$, a chemical potential $2\t \mu$
and a SC gap $\Del(\v k)$. 
The high energy gap closes and spectral weight on one branch vanishes,
yielding a BCS spectrum as shown in Fig. 4(c).

We have also calculated the effect of constant $\v A$ on the qp
dispersion, to linear order of $\v A$. This adds a term 
$\frac1c \v j_\pm \cdot \v A$ to Eq. (14) where 
$\v j_\pm $ is interpreted as the current carried by the qp. We
recall that in standard BCS theory, the current is given in term of the normal
state spectrum by $c \prt_{\v A} \eps_{\v A}=e \prt_{\v k} \eps$ because 
$\eps_{\v A} (\v k) = \eps(\v k+ \frac{e}{c} \v A)$. 
Remarkably this is almost true
in our case in the sense that $\v j_\pm$ is given by 
$c \prt_{\v A} \t E_{\pm, \v A}$, where $\t E_{\pm, \v A}$ 
is obtained by replacing
$\v k$ by $\v k + \frac{e}{c} \v A$ 
in $\eps$, $\t \mu$ and $\Del$ everywhere  in
Eq.\ref{mc17} except for the term $\left( \frac{x_{pc}}{x} \Del \right)^2$, which
is kept independent of $\v A$. Near the node, $\Del$ is negligible 
so that the current is very close to $e \prt_{\v k} \t E \simeq 
e \prt_{\v k} E^N$
(which becomes exactly $e\prt_{\v k} \eps$ along the diagonal), 
thus reproducing Eq. \ref{Esc}. 
We have checked numerically that
even away from the node in the region of the ``FS segment'',
the current is remarkably close to $e \prt_{\v k} E^N$,
which can be quite different from the BCS value 
$e \prt_{\v k} \eps$ near the edge of the FS segment.

From Eq. (6), the temperature dependence of the London
penetration depth gives a direct measurement of $\alpha^{2} \frac{v_F}{v_2}$. 
Density of
states measurements using the $T^2$ coefficient of the specific heat yields
$v_F v_2$. The Fermi velocity can be estimated from transport 
measurements  or
high resolution photoemission experiment. Thus in principle the quantities
$\alpha$, $v_F$ and $v_2$ can be measured. It is of course of great interest to
establish how close $\alpha$ is to $1$, or whether $v_2$ is reduced with respect
to that extrapolated from the energy gap at $(0,\pi)$ measured by
photoemission or tunneling. Crude estimates made in Ref. \cite{LeeWen} suggest
that $\alpha$ is consistent with $1$ but a more precise measurement is clearly
called for.

Finally we comment on finite temperature behaviors. In addition to the
reduction of superfluid density due to thermal excitation of qp, 
 we expect $x_{pc}$ to
decrease with increasing $T$, leading to a reduction of $v_2$:
$v_2(T)=\frac{x_{pc}(T)}{x_{pc}(0)}v_2(0)$.  As $T$ reaches $T_c$, $x_{pc}=v_2=0$ and the nodes of
$E^{(sc)}$ become the ``FS segment'' while the spin gap near $(0,\pi)$ remain finite. We
see that $x_{pc}$ plays the role of the order parameter of the transition,
so that
we may expect the temperature dependence of $x_{pc}$
to be described by a Ginzburg-Landau theory with X-Y symmetry near the
transition. 

\section{Conclusions and Open Issues}
We believe the $SU(2)$ slave boson theory captures the basic physics of the underdoped
cuprates.  The many anomalous properties associated with the spin gap formation are explained
in a natural way.  Superconductivity with $d$-wave pairing symmetry emerges naturally, with
quasiparticle excitations which are remarkabley similar to BCS theory.  However, the
microscopic mechanism is completely different in that the SC state is not formed out of
pairing of normal state quasiparticles via exchange of some effective interaction.  Instead,
it is the coherence of the charge degrees of freedom which converts the spin gap phase to the
SC state.  Many open issues remain, however, and we list a few of them below.

\begin{itemize}
\item[1)] Our discussion of the electron spectrum in the normal state is still at a crude
level.  We treat the bosons as ``nearly'' bose condensed with a relatively narrow spectral
function.  Thus we do not have a theory of the lineshape.  One of the most important
features of the photoemission experiment is that a narrow qp peak forms out of a broad
lineshape as the SC state develops out of the normal state.  We are unable to describe this
evolution at present.  A narrow spectral line is very natural in the single boson
condensation scenario but not as obvious in the boson pair condensation scenario. Thus we
have not achieved a quantitative description of the recombination of spin and charge to form
quasiparticles in the superconducting state.  A related
issue is that in our theory the spin gap state and the SC state share the same
energy scale, i.e., the energy gap $\Delta_0$ at $(0,\pi)$.  Empirically $\Delta_0 \approx
J/3$, in rough agreement with the gap calculated in mean field theory.  Recently, Shen and
collaborators \cite{Ronning} have focused on a higher energy scale (of order $J$ to $2J$)
which characterizes the location of the peak in the ARPES spectrum, and argued that it is the
peak energy which is smoothly connected with the insulator at half-filling.  In this scenario
one would need a separate mechanism to produce the leading edge shift and the SC energy gap. 
In our scenario we have only one energy scale $\Delta_0$ and the burden upon us is to show
that the lineshape may exhibit a peak at high energy of order $J$.

\item[2)] We do not have a satisfactory theory of the transport of the normal state.  This is
related to the still lack of understanding of how the spin-charge separation state in the
normal state evolves to the well defined qp in the SC.  We can only provide a
phenomenological picture of gradual binding between holons and spinons to form physical holes
as the temperature is decreased.\cite{dkkLee}  

\item[3)] The mean field theory underestimates the spin fluctuation near $(\pi,\pi)$. 
While inclusion of gauge fluctuations leads to a satisfactory fit of the specific heat and
uniform spin susceptibility, \cite{Ubbens} it is expected \cite{KimLee} that gauge
fluctuations will strongly enhance the spin fluctuation near $(\pi,\pi)$ but detailed
calculations have not been carried out.  This strong enhancement is needed to explain the
strong peak in the Cu NMR relaxation at a temperature $T^*$ which is low compared with the
spin gap energy $\Delta_0$.  As an intermediate step, we recently carried out a RPA
calculation of the spin fluctuation near
$(\pi,\pi)$.\cite{Brinckmann}  By tuning a single parameter (the effective exchange coupling
in RPA) we are able to account for the resonance peak seen in neutron scattering in the SC
state and its evolution with reduced doping.\cite{Fong}  However, at present we cannot
explain the neutron scattering and the copper NMR within the same RPA theory.
\end{itemize}

The work reviewed in this paper has been done in close collaboration with X.G. Wen and I have
benefitted from collaboration over the years with N. Nagaosa, T.K. Ng, Derek K.K. Lee, 
Don H. Kim, and J. Brinckmann.  This work was supported by NSF through the MRSEC program DMR
98--08941.

\end{document}